\documentclass{article}

\usepackage{arxiv}

\usepackage[utf8]{inputenc} 
\usepackage[T1]{fontenc}    
\usepackage{hyperref}       
\usepackage{url}            
\usepackage{booktabs}       
\usepackage{amsfonts}       
\usepackage{nicefrac}       
\usepackage{microtype}      
\usepackage{lipsum}		
\usepackage{amsthm,amsmath}
\usepackage{graphicx}
\usepackage[sort, numbers]{natbib}
\usepackage{doi}
\usepackage[utf8]{inputenc} 
\usepackage{textgreek} 
\usepackage{bm} 
\usepackage{graphicx}
\usepackage[separate-uncertainty=true,multi-part-units=single]{siunitx}

\usepackage[capitalise]{cleveref}
\usepackage{csquotes}
\usepackage{booktabs}
\DeclareSIUnit\bar{bar}
\DeclareUnicodeCharacter{2060}{\nolinebreak}

\title{Demonstration of tritium adsorption on graphene}

\author{Genrich Zeller
    \thanks{Tritium Laboratory Karlsruhe (TLK), Institute for Astroparticle Physics (IAP), Karlsruhe Institute of Technology (KIT), Hermann-von-Helmholtz-Platz 1, 76344 Eggenstein-Leopoldshafen, Germany}\\ 
	\texttt{genrich.zeller@kit.edu} 
	\And
	Desedea D{\'{\i}}az Barrero$^*$
    \thanks{Departamento de Química Física Aplicada, Universidad Autónoma de Madrid, Campus de Cantoblanco, 28049 Madrid, Spain}\\ 
    \And
    Paul Wiesen$^*$
    \And
    Simon Niemes$^*$
    \And
    Nancy Tuchscherer$^*$
    \And
    Max Aker$^*$
    \And
    Artus M. W. Leonhardt$^*$
    \And
    Jannik Demand$^*$
    \And
    Kathrin Valerius
    \thanks{Institute for Astroparticle Physics (IAP), Karlsruhe Institute of Technology (KIT), Hermann-von-Helmholtz-Platz 1, 76344 Eggenstein-Leopoldshafen, Germany}\\ 
    \And
    Beate Bornschein$^*$
    \And
    Magnus Schlösser$^*$
    \And
    Helmut H. Telle$^\dagger$ \\
}

\renewcommand{\shorttitle}{Demonstration of tritium adsorption on graphene}

\begin{document}
\maketitle
\begin{abstract} 
\large
    In this work, we report on studies of graphene exposed to tritium gas in a controlled environment. 
    The single layer graphene on a $\textrm{SiO}_2$/Si substrate was exposed to $\SI{400}{\milli\bar}$ of $\textrm{T}_2$ for a total time of $\approx \SI{55}{\hour}$. 
    The resistivity of the graphene sample was measured \textit{in situ} during tritium exposure using the Van der Pauw method. 
    We found that the sheet resistance increases by three orders of magnitude during the exposure, suggesting significant chemisorption of tritium. After exposure, the samples were characterised \textit{ex situ} via spatio-chemical mapping with a confocal Raman microscope, to study the effect of tritium on the graphene structure (tritiation yielding T-graphene), as well as the homogeneity of modifications across the whole area of the graphene film. The Raman spectra after tritium exposure were comparable to previously observed results in hydrogen-loading experiments, carried out by other groups. By thermal annealing we also could demonstrate, using Raman spectral analysis, that the structural changes were largely reversible. Considering all observations, we conclude that the graphene film was at least partially tritiated during the tritium exposure, and that the graphene film by and large withstands the bombardment by electrons from the \textbeta-decay of tritium, as well as by energetic primary and secondary ions.
\end{abstract}


\clearpage

\newcommand{\todo}[1]{{\color{red}{#1}}} 
\newcommand{\comment}[1]{{\color{blue}{#1}}} 

\newcommand{\myparagraph}[1]{\paragraph{#1}\mbox{}\\}

\large
\section{Introduction}
Graphene – a single layer of carbon atoms arranged in a two-dimensional honeycomb lattice – has captured the attention of scientists,
engineers, and innovators worldwide, due to its extraordinary properties since its discovery. \cite{Novoselov.2004Electricfieldeffect}. 

Among its many potential applications, one of the most promising is its use in hydrogen storage and utilization, for which its interactions must be well known 
\cite{CastellanosGomez.2012Reversiblehydrogenationand, Dzhurakhalov.2011Structureandenergetics, Elias.2009Controlofgraphene, Fei.2020Synthesispropertiesand, Felten.2014Insightintohydrogenationb}.⁠
Now, studies for potential applications are also extended to the other hydrogen isotopes, i.e., deuterium \cite{Abdelnabi.2021DeuteriumAdsorptionon,Bocquet2014-pa} and tritium \cite{Zhang.2022AdsorptionandDesorption,Rethinasabapathy}.⁠

Tritium is the fuel for future fusion reactors and is also
present as by-product in fission power plants. In this context, the graphene-tritium system is studied \cite{Zhang.2022AdsorptionandDesorption},⁠ and its properties are considered for tritium-processing applications \cite{LozadaHidalgo.2017Scalableandefficient}.⁠ 

The motivation for the research presented in this paper, however, stems largely from the field of astroparticle physics. Current-generation neutrino mass experiments, like KATRIN \cite{Aker.2022KATRIN:statusand, Aker.2022Directneutrinomassmeasurement}, are limited in sensitivity not only by statistics, but also from the molecular nature of tritium in the \textbeta-electron source \cite{Bodine.2015Assessmentofmolecularc}
\begin{equation}
    \textrm{T}_2 \longrightarrow ({}^3\textrm{HeT})^+ + e^- + \bar{\nu}_e. \label{betadecay}
\end{equation}

The $({}^3\textrm{HeT})^+$ molecular ion ends up in a distribution of electronic, vibrational, and rotational states; its final-state distribution (FSD) leads to an effective energy broadening of the spectrum of about \SI{0.4}{\electronvolt} \cite{Bodine.2015Assessmentofmolecularc}, which is limiting the neutrino mass sensitivity to about \SI{0.1}{\electronvolt\per c\squared}. 

In order to avoid this molecular broadening in the \textbeta-decay one viable option is to use an atomic tritium source,
\begin{equation}
    \textrm{T} \longrightarrow ({}^3\textrm{He})^+ + e^- + \bar{\nu}_e. \label{atomic-betadecay}
\end{equation}

In this context, an atomic tritium source is key to the experiment undertaken by the \textit{Project8} collaboration \cite{Project_8_Collaboration2022-mw}, and is considered for future stages of KATRIN successor experiments. In brief, the proposed T-atom souce is based on thermal dissociation of molecular T$_2$, followed by several cooling steps in an atomic beam, and finally trapping T-atoms in a magnetic trap. Proof-of-concept studies are under way, at present using hydrogen as a testbed platform.

Another approach for determining the electron neutrino mass was proposed by the \textit{PTOLEMY} collaboration \cite{Baracchini.06.08.2018PTOLEMY:AProposal}; said experiment is designed to study the cosmic neutrino background by inverse \textbeta-decay\cite{Weinberg.1962UniversalNeutrinoDegeneracy}, 
\begin{equation}
    \nu_e + \textrm{T} \longrightarrow ({}^3\textrm{He})^+ + e^-. \label{inverse-betadecay}
\end{equation}

For this experiment, the intriguing concept of using tritium bound on graphene was suggested, to serve as a quasi-atomic, solid-state tritium target. In the proposal it was postulated that the aforementioned final-state distribution would play a significantly lesser role in comparison to molecular T$_2$ \cite{Betts.17.07.2013Developmentofab}. However, two potential obstacles can be identified for such a tritium source/target. First, recently it was argued that some energy spread of the emitted \textbeta-electron will inevitably be encountered after it is generated in the decay of tritium bound to graphene. \cite{Cheipesh.2021Navigatingthepitfalls,Nussinov.2022Quantuminducedbroadening} Second, the required large-scale T-graphene target poses a significant technical challenge, and its fabrication is still unproven. Meanwhile, different carbon-based substrates are being considered, such as carbon-nanotubes or free-standing nano-porous graphene. \cite{Abdelnabi.2021Towardsfreestandinggraphane}.

Therefore, in order to judge the applicability of large-scale tritium/carbon systems, we believe it is imperative to investigate their fundamental properties. Tritium is well-known for tits aggresive radiochemical nature, which could well make the formation of stable, tritiated structures a great challenge.

The goal of the work presented here was to chemisorb tritium on a graphene-monolayer, on a SiO$_2$/Si substrate. Hydrogenation of graphene is usually performed with thermal molecule-crackers (generating atomic hydrogen), or with plasma sources (generating atomic and ionic hydrogen) \cite{Elias.2009Controlofgraphene,Whitener.2018ReviewArticle:Hydrogenatedb, Sofo.2007Graphane:Atwodimensionalb}⁠. 
Building analogous, tritium-compatible equipment operated in a licensed laboratory is expensive, laborious, and time-consuming. Thus, prior to this step, in this work we followed a different approach, namely, to generate the tritium atoms /ions `naturally' via self-radiolysis, as a consequence of \textbeta-decay and sub-sequent ionizations / dissociations withing the tritium gas environment. In addition, due to the radioactive and volatile properties of tritium, many considerations need to be made regarding legal regulations, safety, and equipment contamination.

Furthermore, those constraints severely limit the choice of possible characterization measurements. In this work, we employ sheet resistance measurements for \textit{in situ} monitoring, using the Van der Pauw method \cite{L.J.vanderPauw.1958AMethodof}. Resistance measurements are commonly used to monitor, and quantifiy, changes in the hydrogenation level of graphene. \cite{JonathanGuillemette.01.2014ElectronicTransportin, Son.2016Hydrogenatedmonolayergraphene}
For \textit{ex situ} characterization measurements, Raman spectroscopy was chosen, which has been proven to be a versatile tool for graphene studies. \cite{Son.2016Hydrogenatedmonolayergraphene, Beams.2015Ramancharacterizationof, Casiraghi.2009Probingdisorderand, Ferrari.2013Ramanspectroscopyas, Lucchese.2010Quantifyingioninduceddefects, Pollard.2014Quantitativecharacterizationof, Eckmann.2012Probingthenature} Here we used a confocal Raman microscope (CRM) which was designed and built specifically for radioactive (or toxic) samples. \cite{DiazBarrero.2022VersatileConfocalRaman}

Using two or more complementary characterization methods is crucial to gain an understanding about the nature of the tritium-graphene interaction. Since this is the first time the effect of tritium on graphene is measured, no possible outcome can be excluded beforehand. For example, tritium could react with the carbon from graphene to tritiated methane \cite{DiazBarrero.2023GenerationandAnalysis}, thus damaging or destroying the graphene layer. It is therefore necessary to distinguish between vacancy-type and sp$^3$-type defects. For this, additional measurement methodologies have been applied to assist in the interpretation of the Raman data (see Section 2.4).
\section{Experimental section}
\subsection{Graphene samples}
The graphene samples employed in this work are monolayer graphene on \SI{90}{\nano\meter} SiO$_2$/Si-substrates (<100> Si mono-crystals of thickness \SI{525}{\micro\meter}, with \SI{90}{\nano\meter} SiO$_2$ coatings on both sides). The graphene samples are $\SI{1}{\centi\meter}\times \SI{1}{\centi\meter}$ in size, and according to the manufacturer (\textit{Graphenea}, San Sebastián, Spain) the graphene film has a sheet resistance of $R_S = \SI{350\pm 40}{\Omega\per}\square$;~\cite{Graphenea.MonolayerGrapheneon} the symbol~$\square$ stands for the total sheet area.
Note that the density of carbon atoms on a graphene surface is about \SI{3.86e19}{atoms \per \meter\squared}~\cite{Pop.2012Thermalpropertiesof}, thus the \SI{1}{\centi\meter\squared} graphene layer corresponds to $N_C = \num{3.86e15}$ carbon atoms. 

\subsection{Setup for exposing samples to tritium}
\begin{figure}[htp]    
    \centering
    \includegraphics[width=.45\textwidth]{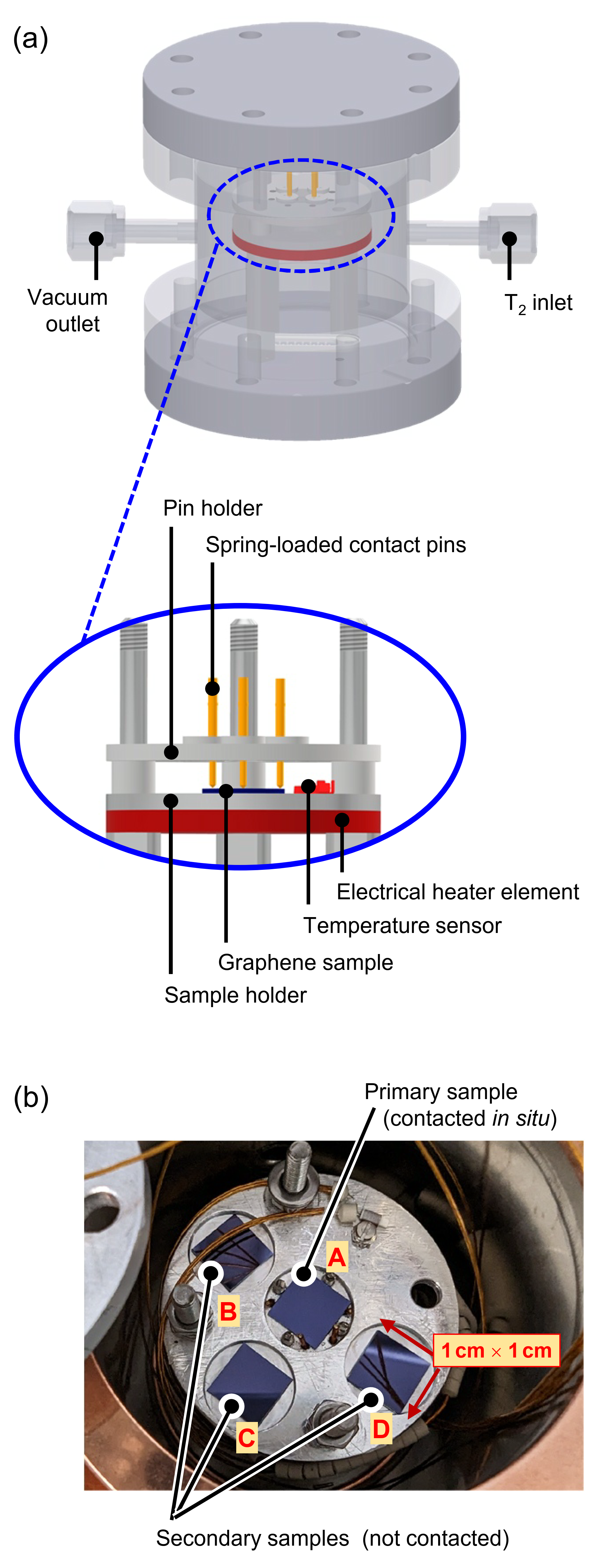}
    \caption{ Experimental setup of the tritium loading-chamber. (a) 3D-view of technical drawing of the loading chamber, and cross-section view of the contacted and heated sample holder. (b) Sample holder with 4 graphene samples, one electrically contacted (centre) and three without contacts. 
    }
    \label{fig:experimental-setup}
\end{figure}

A custom-made, proto-type loading chamber was used for tritium exposure of the graphene samples; the principal constraction layout of the loading chamber is shown in \cref{fig:experimental-setup}, with key components indicated. 

The stacking design allows for easy handling within a glovebox; all components are fully tritium compatible and are made from suitable materials, like stainless steel, aluminium, copper, and ceramics. 
Four graphene samples are place in close proximity on a sample holder (see \cref{fig:experimental-setup}b9 and are exposed to tritium simultaneously. 

The center (\textit{primary}) sample is contacted directly via four spring-loaded contacts (\textit{PTR Hartmann}, Werne, Germany), which are used for the measurent of the graphene sheet resistance via the Van der Pauw method \cite{L.J.vanderPauw.1958AMethodof}. The sheet resistance measurements are conducted using a \textit{DAQ6150} with a \textit{7709-matrix switching card} (both from \textit{Keithley}, Cleveland, USA). 
In order to characterize the temperature dependence of the graphene sheet resistance, an electrical heater (\textit{Thermocoax}, Heidelberg, Germany) and a temperature sensor (\textit{Allectra}, Schönfliess b. Berlin, Germany) are installed close to the sample. 

The \textit{secondary} samples can be used for, e.g., the (destructive) activity determination measurements to assess the adsorbed activity before further handling of the samples.

\subsection{Cold tritium plasma}
In general, for the hydrogenation of graphene, either the hydrogen of the graphene has to be chemically activated.\cite{Whitener.2018ReviewArticle:Hydrogenatedb} In contrast, for the tritiation process utilised during this study self-radiolysis of tritium has been taken advantage of.
The tritium loading gas mixture was provided by the TRIHYDE facility\cite{Niemes.2021Calibrationofa, Niemes.2021AccurateReferenceGas} of the Tritium Laboratory Karlsruhe (TLK) and consisted of \SI{97.2}{\percent} T$_2$ with the remaining \SI{2.8}{\percent} being mainly HT and DT. The pressure in the loading chamber was about \SI{400}{\milli\bar} during the exposure. Given the chamber volume of about \SI{0.2}{\liter}, this corresponds to a total activity of \SI{7.6e12}{\becquerel}.

Hydrogen atoms and ions can either reflect from, adsorb to, or penetrate the graphene lattice. 
Most studies describe atom $\leftrightarrow$ graphene interactions, but Despiau-Pujo et al.\cite{DespiauPujo.2016HydrogenPlasmasProcessing} argued that the energy ranges governing the graphene surface interactions are similar for atoms and ions, and thus ions could contribute to the chemisorption process, in principle.
For tritium, Nakamura et al.\cite{Nakamura.2008MolecularDynamicsSimulation} and later Wu et al. \cite{Wu.2022Adsorptionofhydrogen} calculated that a significant adsorption probability for tritium atoms of \SIrange{25}{75}{\percent}( $\bar{p} \approx\SI{50}{\percent}$) can be achieved for kinetic energies between \num{0.4} and \SI{10}{\electronvolt}. 

In addition to these non-destructive processes, (partial) destruction of the graphene surface must be considered as well; this always becomes a possibility on exposure to a hydrogen plasma. While in general such (irreversible) destruction is not desirable, Despiau-Pujo and co-workers\cite{DespiauPujo.2016HydrogenPlasmasProcessing} discussed in their publication how one might exploit H-plasma interaction to clean, fictionalize and pattern (i.e., tailor the structure and the respective properties) graphene layers in a controlled way.

According to theory \cite{Saenz.2000ImprovedMolecularFinalState,Jonsell.1999Neutrinomassdeterminationfrom}, and as recently experimentally verified \cite{Lin.2020BetaDecayof}, about half of the \textbeta-decays of T$_2$ lead to a bound state of HeT$^+$ (see Eq. 1), while the other half yields the dissociation products He + T$^+$ or He$^+$ + T afterwards \cite{Bodine.2015Assessmentofmolecularc} with efficiency $\eta_{\textrm{dec,diss}} \approx \SI{50}{\percent}$.\cite{Baracchini.06.08.2018PTOLEMY:AProposal}. The dissociation products exhibit kinetic energies in the range \SIrange{3}{13}{\electronvolt}; \cite{Baracchini.06.08.2018PTOLEMY:AProposal} they quickly thermalize down to tens of meV by collisions with the gas, which is at room temperature. For initial particle energies of O(\SI{1}{\electronvolt}) the inelastic scattering cross-section (of H$_2^+$ or H$_2$) is about \SI{1e-15}{\centi\meter\squared} \cite{tabata.2000ANALYTICCROSSSECTIONS}. This equates to a mean free path of about one micrometre (\SI{1}{\micro\meter}). 

When T / T$^+$ interact with the T$_2$-gas, further processes take place, \cite{Klein.2019Tritiumionsin} such as ionization, gas phase formation of larger ion clusters (T$^+ ~\rightarrow$ T$_3^+~\rightarrow$ T$_5^+ ~\rightarrow$ ...), and the recombination of the ions with electrons. 
The secondary molecular ions dissociate, with an efficiency of $\eta_{\textrm{scat,diss}} \approx \SI{5}{\percent}$. 
The resulting T$_n^+$ species receive between \SI{0}{\electronvolt} and \SI{15}{\electronvolt} of kinetic energy, peaking at about \SI{8}{\electronvolt} \cite{Dunn.1963DissociativeIonizationof}. The calculation of the rate of ion generation with respect to initial \textbeta-decay electrons is complex but can be obtained, in principle, by Monte-Carlo methods \cite{Kellerer.2022MonteCarlosimulations}.

Note that most scattering partners (T$_2$ vs T, T$^+$, T$_2^+$, T$_3^+$, ...) are of similar mass; thus in every collision the particle loses on average less than \SI{50}{\percent} of its kinetic energy: about four to five scattering steps are required to drop the kinetic energy below the threshold for adsorption. 
Based on this, the volume above the \SI{1}{\centi\meter\squared} graphene sample, in which atoms / ions are generated with sufficiently low energy for tritium chemisorption at the surface, is estimated to be about $\SI{1}{\centi\meter} \times \SI{1}{\centi\meter} \times (5\times 1)\si{\micro\meter}$.

Here we like to point out that, at present we do not report accurate calculations of ion / atom fluxes onto the graphene film. We only estimate that in principle tritium ions / atoms have been produced in sufficient quantity in the energy range of interest, to provide significant tritium adsorption during the exposure time. At the same time, one will encounter a fraction of the ions / atoms whose energies still remain sufficiently high to be able to introduce damages to the graphene layer.

\subsection{Measurement methodology}
Due to the radioactive and volatile nature of tritium, many considerations must be made regarding regulatory requirements, safety, and contamination of the equipment. 
This severely limits the choice of possible characterization measurements. In contrast to most experiments with hydrogen, most steps of a tritium experiment are time-consuming and laborious. 
For example, in order to extract the samples from a tritium loading chamber, the chamber needs to be evacuated for at least a few days to minimize contamination of the surrounding glove box.

Therefore, for experiments with tritium, it is essential to
incorporate at least one \textit{in situ} characterization method, besides a range of \textit{ex situ} analysis tools.

In this work, \textit{in situ}, real-time sheet resistance monitoring was utilized, and \textit{ex situ} Raman characterization measurements, in combination with thermal annealing in a tritium-compatible oven, were used to investigate the nature of graphene defects introduced by its exposure to tritium. Finally, total sample activity determination helped in the evaluation of the actual tritium coverage; unfortunately, this latter measurement is destructive.
\subsubsection*{\textit{In situ} sheet resistance measurements}
One simple method for \textit{in situ} monitoring is the sheet resistance measurement of graphene.\cite{JonathanGuillemette.01.2014ElectronicTransportin} By using a four-point-resistance measurement via the Van der Pauw method,\cite{L.J.vanderPauw.1958AMethodof} the sheet resistance of the graphene sample can be measured offset-free and compared to similar experiments by other groups.
However, temperature-dependence measurements - similar to those reported for hydrogenated graphene\cite{Son.2016Hydrogenatedmonolayergraphene} - are not included here, because our Van der Pauw measurement setup failed for temperatures above \SI{120}{\degree C}, probably related to thermal stress in contacts between graphene and the spring-loaded electrodes. For additional details, see Supplementary Information S2.

Also, it has to be stressed that using sheet resistance measurements alone, it is not possible to distinguish between the types of defects introduced to the graphene layer. Two main types of defects in graphene are relevant to this work, namely vacancy- and sp$^3$-type defects. In the literature, three main methods are employed to distinguish between these defect types, as outlined below.

\subsubsection*{\textit{Ex situ} X-ray photoelectron spectroscopy (XPS)}
X-ray photoelectron spectroscopy can be used to measure the bond energy directly and is therefore the method of choice when available.\cite{Speranza07}
XPS systems are expensive and therefore not favourable for the use with radioactive tritium samples, which could experience out-gassing of tritium. For this work, no XPS system was available; however, XPS data from the literature for exposure to hydrogen were used,\cite{Son.2016Hydrogenatedmonolayergraphene} for cross-comparison of the sheet resistance and Raman measurements (see Section 3.1).
\subsubsection*{\textit{Ex situ} Raman spectroscopy} 
While ideally \textit{in situ} Raman spectral monitoring during the tritium exposure of graphene would be incorporated, this was not possible, due to no optical access possibility in the very basic proof-of-concept construction of our loading chamber.
Thus, \textit{ex situ} (pre- and post-exposure with tritium) Raman spectra were collected using a custom-built confocal (imaging) Raman microscope; \cite{DiazBarrero.2022VersatileConfocalRaman} for completeness, the conceptual setup of our CRM is summarised in Supplementary Information S3. The microscope was equipped with a 10x objective lens (NA = 0.25), resulting in a laser focal beam diameter (FBD) on the graphene surface of FBD $\approx\SI{7.3}{\micro\meter}$.

All Raman measurements were carried out using a \SI{532}{\nano\meter} excitation laser, with a laser power of \SI{120}{\milli\watt} (power density on the graphene surface $\approx \SI{3e5}{\watt\per\centi\meter\squared}$). Even after prolonged exposure of several minutes at this power density, we did not observe changes in or damage of the graphene sheet. For the determination of the peak-intensities and line-widths, the respective Raman peaks are fitted with a Lorentzian function during spectral data analysis.

\subsubsection*{\textit{Ex situ} thermal annealing}
By thermal annealing it can be investigated, whether the observed changes to the graphene layer are reversible. \cite{Son.2016Hydrogenatedmonolayergraphene}
Although graphene possesses self-healing properties, \cite{Chen.2013Selfhealingof} severe damages (vacancy-type defects) are only completely reversible in the presence of hydrocarbon gases.\cite{Lopez.2009ChemicalVaporDeposition}
Several studies have demonstrated the de-hydrogenation of graphene at temperatures above \SI{300}{\degree C}. \cite{Son.2016Hydrogenatedmonolayergraphene,Cha.2022Damagefreehydrogenationof} Thus, by heating the tritium-exposed samples, it can be ascertained, whether effects caused by tritium exposure are reversible; if reversibility were found, this would strongly suggest sp$^3$-type C-T bonding.

\subsubsection*{\textit{Ex situ total activity determination}}
Using a tritium compatible oven, the graphene samples can be heated \textit{ex situ} to up to \SI{1600}{\degree}C, in an oxygen-containing gas stream; this severe heating removes all tritium from the sample. However, at the some time the graphene layer is destroyed as well.

The released activity is measured using a proven TLK setup. In short, the exhaust from the oven - mostly in the form of T$_2$ and HTO - passes through an oxidising CuO-wire bed, and then through a water bubbler, where all tritiated species are retained. The content of the water bubbler is then used to determine the total activity released during the sample heating, via liquid scintillation counting. This can also provide additional information about the nature of the C$\leftrightarrow$T interaction.

\begin{figure}[tbp]
    \centering
    \includegraphics[width=.95\textwidth]{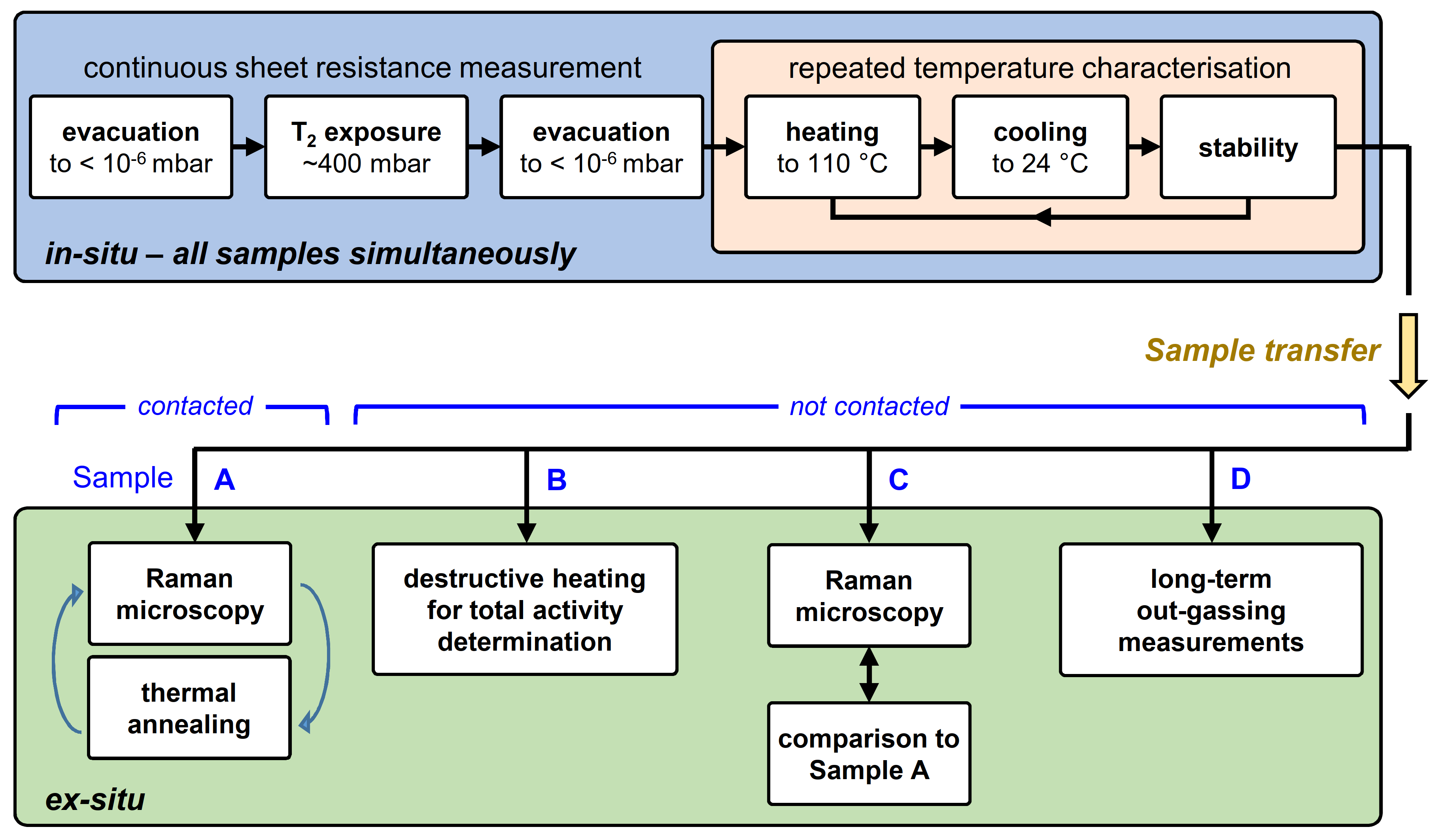}
    \caption{Measurement methodology for tritium-exposed graphene samples. Top - exposure of sample to tritium, followed by repeated heating cycles (monitored \textit{in situ} by resistance measurement); bottom - \textit{ex situ }characterisation measurements. For details, see text.}
    \label{fig:overview}
\end{figure}

\subsubsection*{Measurement protocol}
The overall measurement methodology can be sub-divided into two main action blocks. These comprise (i) the tritiation process of the graphene samples, including \textit{in situ} stability measurements of the tritiated samples; and (ii) \textit{ex situ} T-graphene characterisation after termination of the exposure; this in shown in form of a chart diagram in \cref{fig:overview}. Details for the individual steps are provided in Supplementary Information S1, together with numerical values of key operational parameters.

\section{Results and discussion}
In the following presentation of results and their discussion, we are guided by a particular hypothesis, and we are looking into observations in its favour, or against it. The hypothesis is: 
\begin{center}
\textit{“We can tritiate a graphene lattice (i.e., forming C-T bonds) by exposing it for \SI{55}{\hour} to an atmosphere of almost pure T$_2$ gas, at a pressure of \SI{400}{\milli\bar}”.}
\end{center}

\subsection{\textit{In situ} sheet resistance measurements}
\begin{figure}[b!]
    \centering
    \includegraphics[width=.95\textwidth]{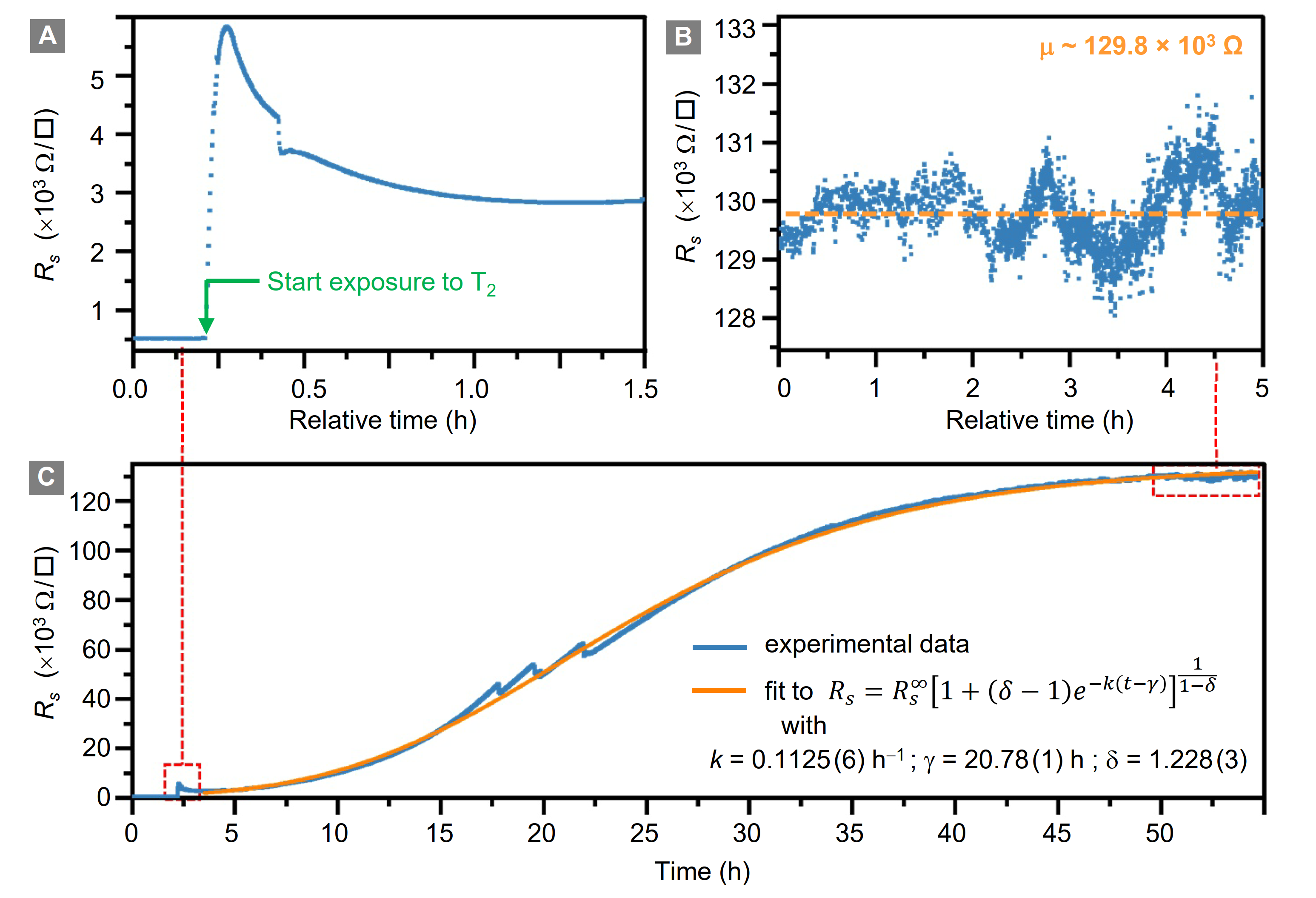}
    \caption{Change of graphene sheet resistance, $R_S$, during tritium exposure. (A) Initial increase of $R_S$ when the loading chamber is filled with tritium. (B) Plateau reached for $R_S$ after \SI{50}{\hour} of tritium exposure. (C) Full temporal evolution of $R_S$ during tritium exposure (orange line = generalised logistic fit to the data).
    }
    \label{fig:loading-curve}
\end{figure}

During this whole tritiation process the sheet resistance,$R_S$, was continuously monitored; these data are shown in \cref{fig:loading-curve} from shortly before the inlet of tritium, and throughout the exposure to tritium at an activity density of \SI{3.4e10}{\becquerel\per\centi\meter\cubed} (at \SI{400}{\milli\bar}). 

In panel (A) the initial increase of the sheet resistance upon inlet of the tritiated gas mixture into the loading chamber is shown. Within just a few min, the sheet resistance of the graphene sheet increases from $R_S = \SI{551\pm 2}{\Omega\per}\square$ to $R_S = \SI{5830\pm 5}{\Omega\per}\square$, reaching a local maximum. In the following \SI{1.5}{\hour}, the sheet resistance initially decreases slightly, but increases again thereafter. 
The increase approximately follows a logistic function.\cite{RICHARDS.1959AFlexibleGrowth}

The data for the complete measurement is shown in \cref{fig:loading-curve}C, together with the fit to the logistic function (orange trace); the expression for the generalised logistic function in included in the figure. the function parameters include the logistic growth rate, $k$; the function's midpoint time, $\gamma$; and a parameter $\delta$ which affects the shape of the growth curve (such as, e.g., the proportion of the final size at which the inflexion point occurs). the numerical values for the associated fit shown in the figure were $k=\SI[uncertainty-mode = compact-marker]{0.1125(6)}{\per\hour}$, $\gamma=\SI[uncertainty-mode = compact-marker]{20.78(1)}{\hour}$, and $\delta=\num[uncertainty-mode = compact-marker]{1.228(3)}$, respectively.

Here we like to point out that the use of generalised logistic functions has been proposed, and this approach is being applied as a common chemical kinetic analysis method. \cite{Avramov2014-tw,Burnham2017-ft}
As such, they describe a behaviour in which a chemical process starts from a base value, increases exponentially and ends in saturation. Indeed, such behaviour is observed in our tritiation experiment, and thus the use of a logistic function to fit the data seems an obvious choice. but one also should keep in mind that at present no complete, analytical model for the plasma evolution of radioactive tritium associated with chemisorption of tritium on graphene exists. Therefore, the description via the logistic function remains phenomenological, and no link between the fit values and the underlying radio-chemical kinetics is immediately obvious.

After an exposure time of about \SI{55}{\hour}, the surface resistance reaches a plateau at about 
$R_S^\infty \approx \SI{120e3}{\Omega\per}\square$,  corresponding to a total relative resistivity increase by a factor of $\approx$~250. This plateau is shown in \cref{fig:loading-curve}B.
It should be noted that the setup used for these measurements can measure resistances up to $\mathcal{O}(\SI{1e6}{\ohm})$, with the measured value well below the instrumental limit.

Son et al.\cite{Son.2016Hydrogenatedmonolayergraphene} cross-calibrated their graphene sheet resistance measurement against XPS-measurements, in which they could quantify the resistivity for two hydrogenation coverage values. 
As mentioned earlier, no XPS system was available to us. While direct comparison between hydrogenation and tritiation data is not possible at present, at least one may arrive at a crude estimate of the coverage. 
Thus, comparing their hydrogenation results (increase of $R_S$ by a factor of 170.9 for $\eta_\textrm{H} = \SI{12}{\percent}$) to our increase in $R_S$, the latter would correspond to a significant tritium coverage of about $\eta_\textrm{T} = \SIrange{10}{20}{\percent}$.

The causes of the decrease of $R_S$ after \SI{5}{\minute} of exposure, as well as the three spikes in $R_S$ visible in \cref{fig:loading-curve}C – after \SI{17}{\hour}, \SI{20}{\hour}, and \SI{22}{\hour} of tritium exposure, respectively – are not yet understood and are subject to ongoing investigations. 

Speculating, the spikes could have been caused by different
effects. These include, for example, vibrations from the surrounding laboratory equipment disturbing the measurement, or a shift of the Fermi level / opening of the bandgap during prolonged tritium exposure, similar to previous results with hydrogen plasmas. \cite{Whitener.2018ReviewArticle:Hydrogenatedb,Son.2016Hydrogenatedmonolayergraphene} Regardless, the general behaviour of $R_S$ is a loading curve comparable to previously reported loading with hydrogen \cite{JonathanGuillemette.01.2014ElectronicTransportin}.

All observations combined clearly demonstrate that there is an interaction between the atoms / ions of the cold tritium plasma and the graphene sheet, which leads to an alteration of the graphene surface, and not to a complete disintegration.

\subsection{\textit{Ex situ} Raman spectroscopy and thermal annealing of tritium exposed samples}
\begin{figure}[tbp]
    \centering
    \includegraphics[width=.95\textwidth]{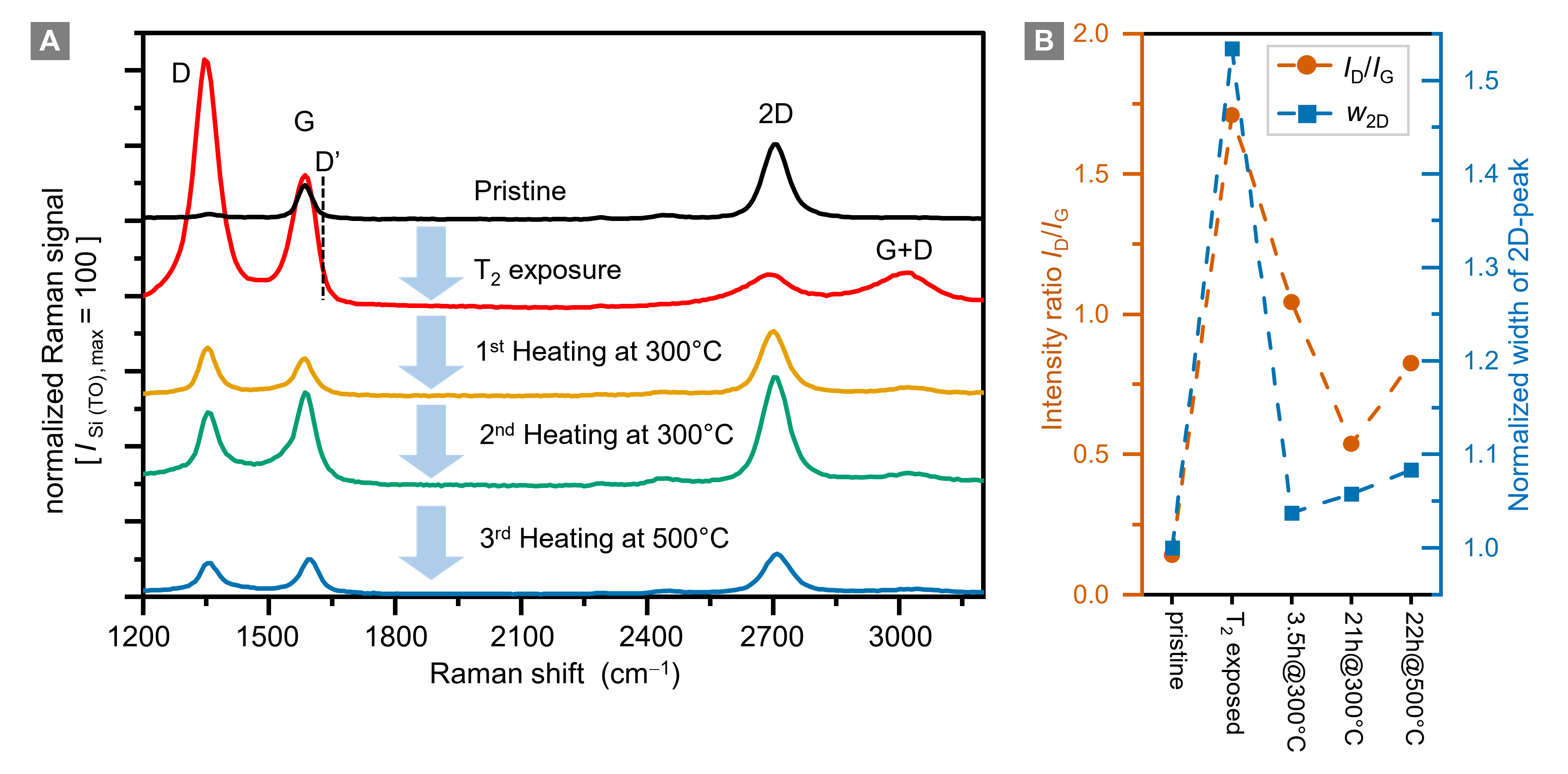}
    \caption{Raman spectra of a graphene sample (A), and intensity ratio $I_\textrm{D}/I_\textrm{G}$ and normalized ($w_\textrm{2D}$-pristine = 1) width of the 2D-peak $w_\textrm{2D}$ (B) - pre-tritium exposure (pristine), post-tritium exposure, and after heating the post-tritium exposure sample 1st for \SI{3.5}{\hour} at \SI{300}{\degree}C, 2nd for \SI{21}{\hour} at \SI{300}{\degree}C, and 3rd for \SI{22}{\hour} at \SI{500}{\degree}C. Raman spectra are shown with a fixed offset for clarity. Key Raman spectral features are annotated.
    }
    \label{fig:raman-spectra}
\end{figure}

Raman spectra of a graphene sample were recorded for five different conditions: (i) before tritium exposure, (ii) after tritium exposure, and (iii-v) after repeated thermal annealing at different temperatures in an Ar-atmosphere. Representative Raman spectra are shown in Fig.~\ref{fig:raman-spectra}A. 

In pristine graphene the dominant features are the Raman G-band ($\sim \SI{1580}{\per\centi\meter}$) 
and the Raman 2D-band ($\sim \SI{2700}{\per\centi\meter}$) \cite{Casiraghi.2009Probingdisorderand}. Both bands are associated with phonon modes without the presence of any kind of defect or disorder \cite{Ferrari.2006Ramanspectrumof}. The intensity ratio $I_\textrm{G}/I_\textrm{2D} < 1$ of the G-peak and the 2D-peak is one indicator for high-quality graphene. 

In defective graphene, several other Raman bands appear in the spectra. For the study of hydrogenated (tritiated) graphene the D-band ($\sim \SI{1340}{\per\centi\meter}$) is the most important feature \cite{Whitener.2018ReviewArticle:Hydrogenatedb}. In addition, the D’-band ($\sim \SI{1620}{\per\centi\meter}$) can be used to distinguish between sp$^3$-type and vacancy-type defects \cite{Eckmann.2012Probingthenature,Eckmann.2013Ramanstudyon}. However, with our current Raman setup, we cannot resolve the D’-band; and it is completely overlapped by the G-band. 

For future investigations, the setup will be upgraded to higher spectral resolution. 
Nevertheless, the small D-peak intensity and the intensity ratio $I_\textrm{D}/I_\textrm{G} < 0.1$ confirm the high-quality of the sample before tritium exposure (see the data in Fig.~\ref{fig:raman-spectra}B). This is also demonstrated by the spatial homogeneity in the Raman map of the sample (see Supplementary Information S4).

In addition, the intensity ratio $I_\textrm{D}/I_\textrm{G}$ is also important because it is related to the defect density on a graphene film. \cite{Lucchese.2010Quantifyingioninduceddefects, MartinsFerreira.2010Evolutionofthe} In this context it should be noted that the D-peak intensity is not monotonic with respect to the defect density. When functionalization levels are very high, the $I_\textrm{D}/I_\textrm{G}$ ratio reaches a maximum and then decreases. In this situation, it is helpful to use other measures to track hydrogenation and dehydrogenation processes.

After exposure to tritium the intensity of the 2D-peak is significantly reduced, while the G-peak intensity increases, resulting in an intensity ratio $I_\textrm{G}/I_\textrm{2D} = 4.8$. In addition, the D-peak intensity increases by a factor of $\sim 70$, becoming the dominant Raman band. The intensity ratio $I_\textrm{D}/I_\textrm{G} = 1.7$ indicates a significant increase in the defect density. In hydrogenated graphene $I_\textrm{D}/I_\textrm{G} = 0.8 \Longleftrightarrow \eta_\textrm{H} = \SI{12}{\percent}$ and $I_\textrm{D}/I_\textrm{G} = 2.18 \Longleftrightarrow \eta_\textrm{H} = \SI{15}{\percent}$ \cite{Son.2016Hydrogenatedmonolayergraphene}, thus implicating - by analogy - a tritium coverage of $\eta_\textrm{T} = \SIrange{12}{14}{\percent}$ for our sample. This is in reasonable agreement with the estimate obtained from the sheet resistance measurement ($\eta_\textrm{T} = \SIrange{10}{20}{\percent}$, see Section~3.1). 

As stated earlier, heating the tritium-exposed samples might provide indications whether changes caused by tritium exposure are reversible. For this, the sample was placed inside a pipe oven at \SI{300}{\degree}C for \SI{3.5}{\hour}. Prior to this 1st heating, the oven was flushed with `wet' Argon gas (flowing through a water wash bottle the Argon gas saturates with water vapor) to prevent oxidation of the graphene layer during heating. 
In this process, heat is transferred from the hot, wet Argon gas to the graphene sample. Here it should be pointed out that the gas and sample temperatures are not measured directly but are lower than the nominal temperature of the ceramic tube of the pipe-oven. Note also that, during this external, thermal annealing of the samples Ar and H$_2$O are present.

While the $I_\textrm{D}/I_\textrm{G}$ ratio decreases from $I_\textrm{D}/I_\textrm{G} =1.7$ to $I_\textrm{D}/I_\textrm{G} = 1.0$, the D-peak does not disappear completely. The 2D-peak intensity is also mostly recovered, reaching $\sim\SI{83}{\percent}$ of the original value of the pristine sample, with an intensity ratio $I_\textrm{G}/I_\textrm{2D} = 0.6$. 
Both observations combined show that the defect density is reduced; in other words, the graphene sample has recovered much of its original properties. 

Cha~et~al.\cite{Cha.2022Damagefreehydrogenationof} have observed a similar, partial reversibility in their hydrogenation experiments, after exposure to a hydrogen plasma with average energies of up to \SI{5.35}{\electronvolt}; they concluded that, the ion energies within the plasma should be between 2.5 and \SI{3.45}{\electronvolt} for damage-free hydrogenation of graphene.  In this context theoretical studies showed that, vacancy defects in graphene form when the energy ranges from \SI{5}{\electronvolt} to \SI{12}{\electronvolt} \cite{Krasheninnikov.2006Bendingtherulesb, ElBarbary.2003Structureandenergetics, Lee.2005Diffusioncoalescenceand,Ertekin.2009Topologicaldescriptionof}. 

In a slightly different approach, Chen et al.\cite{Chen.2013Selfhealingof} have demonstrated `self-healing' of graphene after Ar$^+$-ion bombardment by thermal annealing. In their study, the reduction of the $I_\textrm{D}/I_\textrm{G}$ ratio is even more pronounced, with a minimal value of about $I_\textrm{D}/I_\textrm{G} = 0.25$, after annealing at \SI{800}{\degree}C. However, the relative width of the 2D-peak is increasing significantly (factor > 2) when the annealing temperature exceeds \SI{300}{\degree}C.  This indicates a graphene layer whose quality has worsened. 

 In the experimental section we discussed that the mechanism of producing and cooling the ions down to energies which are compatible with tritium chemisorption onto graphene. Inevitably, a small quantity of ions in the higher energy range are still present. Thus, we expect that the resulting modification of the graphene is in part reversible (T-uptake) and in part irreversible (defect generation). 
 
At similar intensity ratios in comparison to those of Chen~et~al., namely $I_\textrm{D}/I_\textrm{G} \approx 0.5$, we only observe an increase of the 2D-peak width by a factor of $\sim$1.05 after thermal annealing, for a total of \SI{24}{\hour} at \SI{300}{\degree}C (Fig.~\ref{fig:raman-spectra}B). Thus, the quality of the graphene layer is better after the combination of `tritium exposure + thermal annealing' compared to the aforementioned `Ar$^+$-ion bombardment + thermal annealing' of hydrogenated graphene. 
These observations indicate that the change in the Raman spectra seen after thermal annealing exceed the magnitude of the expected effects if self-healing were the only mechanism in play. This supports our hypothesis that we have a significant tritiation effect.

It should also be noted that complete healing of a graphene film was observed in the presence of a hydrocarbon gas,\cite{Lopez.2009ChemicalVaporDeposition}, which however was not present in our annealing oven. In our sample, even after successive thermal annealing for \SI{21}{\hour}, the D-peak remains elevated at $I_\textrm{D}/I_\textrm{G} = 0.53$, suggesting that the quality of the graphene film has decreased permanently. 

In a final, the sample was annealed for \SI{20}{\hour} at \SI{500}{\degree}C. During this process, most of the graphene film was destroyed, and the remaining parts had an increased ratio $I_\textrm{D}/I_\textrm{G} = 0.83$, with increased D-peak intensity. It is therefore clear that with our heating setup graphene is severely damaged at \SI{500}{\degree}C.

\subsection{Total activity on tritium-exposed samples}
As described in the experimental section, the setup for thermal annealing of the samples captures the released tritium while annealing, and the released activity can be quantified using liquid scintillation counting (LSC). The results from the LSC are summarized in Table~\ref{tab:activity}. During each of the three thermal annealing periods of the primary sample, several MBq of activity were released. 

 At this point, it should be stressed that the above indirect methodology for the determination of the sample activity is less than ideal. For quite a few years now beta-induced X-ray spectrometry (BIXS) is being exploited instead for activity monitoring of a gaseous tritium sources or tritium-loaded surfaces; the idea goes back more than two decades.\cite{Matsuyama1998-ct} The method is based on the measurement of characteristic and bremsstrahlung X-rays, induced by \textbeta-electrons from the decay of tritium in the materials. For gaseous samples BIXS measurements are only sensitive to the activity content and are not influenced by the sampling (gas) mixture, as long as the pressure is low enough to avoid significant self-adsorption of the \textbeta-electrons in the sample.

More recently, compact BIXS devices have been designed and tested, that offer convenient integration into any tritium processing/monitoring facility.\cite{Rollig2015-lr} Unfortunately, no such device was available during these tritium-loading experiments, and even if one had been at hand, our rather rudimentary setup would not have had the means to accommodate it. Therefore, the approach outlined earlier in this section had to suffice, in conjunction with literature values for hydrogen coverage. 


\begin{table}[tbp]
\centering
\caption{Activity release from heating of tritium exposed samples} \label{tab:activity}
\begin{tabular}{lllc}
\toprule
Medium & Heating temperature (°C) & Heating duration (h) & \multicolumn{1}{l}{Released activity (Bq) } \\ \midrule
\multicolumn{4}{l}{\textit{Primary sample (contacted)}} \\ \midrule
Ar + H$_2$O & \multicolumn{1}{c}{300} & \multicolumn{1}{c}{3.5} & ($8.0 \pm 1.6$) \\
Ar + H$_2$O & \multicolumn{1}{c}{300} & \multicolumn{1}{c}{21} & ($5.0 \pm 1.0$) \\
Ar + H$_2$O & \multicolumn{1}{c}{500} & \multicolumn{1}{c}{22} & ($6.5 \pm 1.3$) \\
 &  &  & \multicolumn{1}{l}{$\sum = 19.5 \pm 3.9$} \\ \midrule
\multicolumn{1}{l}{\textit{Secondary sample (non-contacted)}} &  &  & \\ \midrule
Air & \multicolumn{1}{c}{1400} & \multicolumn{1}{c}{5} & ($19.0 \pm 4.0$) \\ \bottomrule
\end{tabular}
\end{table}

\subsection{Raman maps}
\begin{figure}[p]
    \centering
    \includegraphics[width=.45\textwidth]{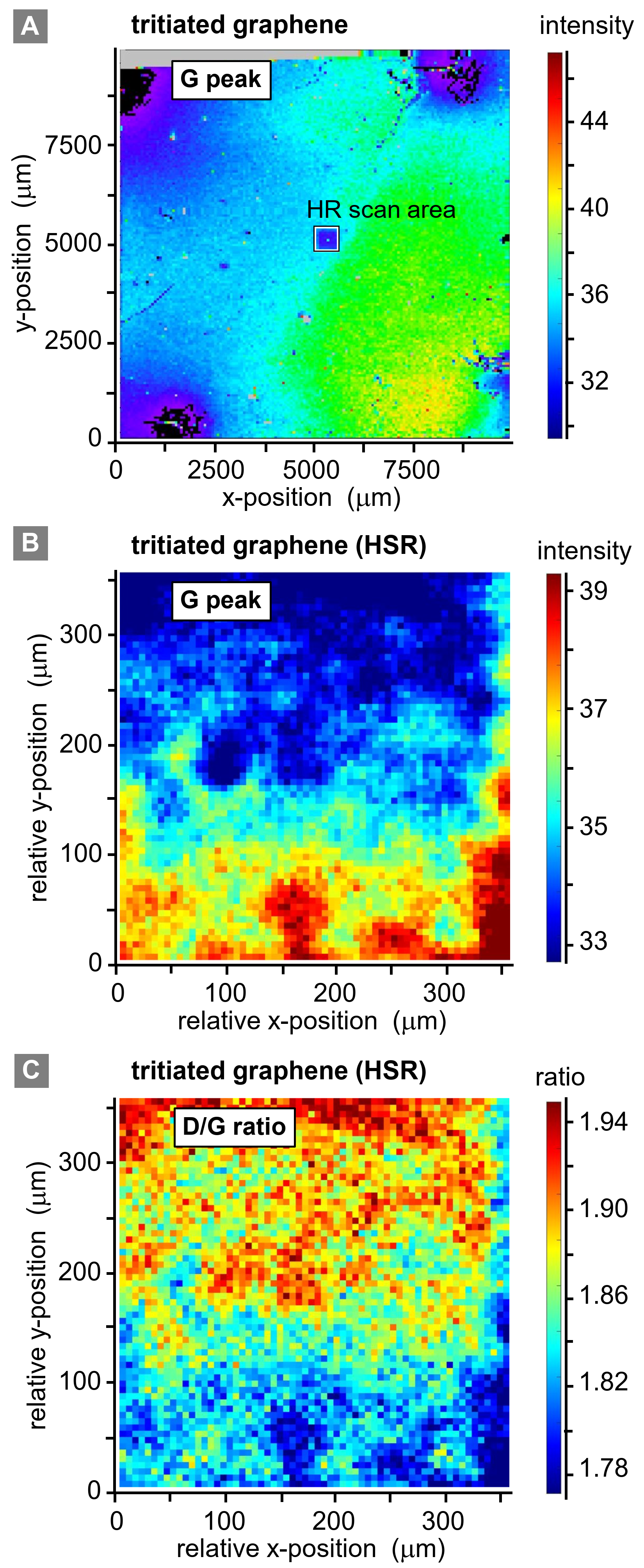}
    \caption{Raman spectroscopic maps of a graphene-on-SiO$_2$/Si sample (\textit{Graphenea}). (A) – Raman map of the graphene G-peak signal, for the full $10 \times \SI{10}{\milli\meter\squared}$ sample post-tritium exposure; step size $\Delta S = \SI{62.5}{\micro\meter}$. (B) Raman map of the graphene G-peak signal, for a $350 \times \SI{350}{\micro\meter\squared}$ sample section post-tritium exposure; step size $\Delta S = \SI{5}{\micro\meter}$. (C) graphene D/G Raman peak ratio map, post-tritium exposure. Note: HSR = high spatial resolution.
    }
    \label{fig:raman-maps}
\end{figure}

Two different lateral scans were conducted on the primary (contacted) tritiated sample: (i) a low spatial resolution (LSR) scan of the full sample ($\SI{1}{\centi\meter}\times\SI{1}{\centi\meter}$) with the step size $\Delta S = \SI{62.5}{\micro\meter}$; and (ii) a high spatial resolution (HSR) scan of a central region on the samples, with the step size $\Delta S = \SI{5}{\micro\meter}$. Different peak and peak-ratio maps from these scans are shown in Fig.~\ref{fig:raman-maps}. 
Equivalent scans of pristine graphene (not shown here) do not add additional information, since with our spatial resolution of $\sim\SI{7}{\micro\meter}$ \cite{DiazBarrero.2022VersatileConfocalRaman}  the pristine samples look very homogenous with a relative standard deviation of the G-peak intensity of only \SI{0.1}{\percent} on an area of $\SI{300}{\micro\meter}\times\SI{300}{\micro\meter}$. However, for completeness they are provided in the Supplementary Information S4.

The LSR scan post-tritium exposure (Fig.~\ref{fig:raman-maps}A) reveals some structures on the scale of up to several \SI{100}{\micro\meter}: 

First, the `black' regions (mainly background signal, associated with sample fluorescence and/or instrument-internal effects) are severely damaged and have nearly no graphene left on them. These regions correspond to the positions of the spring-loaded contact pins. From HSR scans (not included here) it was evident that the contact pins had moved on the graphene surface, either during the initial contacting, vibrations from surrounding vacuum pumps, due to thermal expansion during the \textit{in situ} heating of the sample, or when demounting the sample. 

Second, a radial dependence of the G-peak intensities surrounding the points of contact is observed. This could be caused by shadowing of the main gas volume by the pin-holder, or by some electro-chemical effects induced by the measurement current which is supplied through the contact pins. 

Third, the G-peak intensity is reduced in the region of the HSR scan (which was actually made before the LSR scan); this implies laser-induced or laser-accelerated effects. Their influence is of the order \SIrange{2}{3}{\percent} relative to the initial value; the intensities stabilise after \SI{10}{\hour} of continuous laser exposure. Therefore, this does not significantly affect our working hypothesis of stable chemisorption of tritium.  

Last, the changes associated with the tritium exposure (Fig.~\ref{fig:raman-spectra}A) – represented by the G-peak intensity in Fig.~\ref{fig:raman-maps}A – are distributed smoothly over the sample surface, with a slight spatial gradient. 

From these observations, we conclude that, apart from the points of contact, the graphene film is still covering the whole sample, and there is no large-scale disintegration. The Raman map from the HSR scan (Fig.~\ref{fig:raman-maps}B and C) reveals some substructures on the scale of about $10-\sim\SI{10}{\micro\meter}$, evident by the intensity changes in the map, which could correspond to graphene flake borders. However, overall, the spectral changes are rather moderate (and gradual), with a relative standard deviation of \SI{7}{\percent} for the G-peak intensity and \SI{5}{\percent} for the $I_\textrm{D}/I_\textrm{G}$ ratio.

\section{Conclusions}

 For the first time, graphene was exposed to tritium gas in a controlled environment, with \textit{in situ} real-time monitoring of the graphene sheet resistance, and subsequent post-exposure \textit{ex situ} sample characterisation. These studies confirmed our working hypothesis that the cold-plasma (via self-radiolysis) exposure leads to chemisorption of tritium atoms to the graphene lattice; this is supported by the following findings. 

As a first observation, we report that the sheet resistance develops according to a logistic-growth function during tritium exposure, reaching a plateau after about \SI{55}{\hour}.  In the course of the tritiation process the temperature dependence of the resistance changes, indicating a transition from metallic transport characteristics to insulator-like transport characteristics, as reported in studies with hydrogenated graphene. \cite{Son.2016Hydrogenatedmonolayergraphene} Thus, this strongly indicates chemisorption of tritium to the graphene surface. 

Second, using \textit{ex situ} Raman microscopy, we confirmed that the change in the Raman spectra after tritium exposure is comparable to that observed in hydrogen-loading experiments carried out by other groups. \cite{Son.2016Hydrogenatedmonolayergraphene, Eckmann.2012Probingthenature} Furthermore, the spectral changes are mostly homogeneous, with only slight variations over the whole area of the $\SI{1}{\centi\meter}\times \SI{1}{\centi\meter}$ graphene film.

Third, the Raman spectra recorded after stepwise \textit{ex situ} heating of the samples show that the effect of the tritiation is partially reversible. The 2D-peak and G-peak intensity, and width, can be recovered almost completely, while the D-peak remains at an elevated level resulting in an increased intensity ratio of $I_\textrm{D}/I_\textrm{G} = 0.53$. This suggests that the graphene film was at least partially tritiated (sp$^3$-type associated with C-T bonds). 
At the same time, the elevated D-peak implies defects, which cannot be repaired by thermal annealing (e.g., vacancy-type defects). Therefore, we conclude that, both sp$^3$-type  and vacancy-type defects are present after exposure to tritium, with reversible sp$^3$-type defects being dominant. 

 These observations are compatible with the coarse estimation of possible tritium chemisorption via a mechanism in which atoms and ions of eV-scale energy are generated, likely by dissociation after secondary ionization in collision with \textbeta-electrons, followed by collisional cooling in the gas over a distance of just a few micrometres.

Overall, we have demonstrated that, our rather simple experimental arrangement allows for significant tritiation of a macroscopic graphene surface, and thereby proving the initial working hypothesis to be correct.

While recent theoretical considerations suggest that tritiated graphene may not present a way out of the energy broadening from molecular effects in the \textbeta-electron spectrum, tritium – which is immobilised onto a surface and stable at room temperature – may still offer many practical benefits. In particular, it allows for the preparation of solid-state tritium sources, which may facilitate proof-of-principle studies of modern electron detection concepts. 

Damages introduced to graphene are \textit{per se} not avoidable using our self-radiolysis cold-plasma approach. Other techniques, providing atoms and ions by thermal dissociation or RF plasma sources, might provide more controllable particle energies, and thus be gentler with regard to potential surface damage. Furthermore, any large-scale carbon–tritium electron source, as for example planned for PTOLEMY, \cite{Baracchini.06.08.2018PTOLEMY:AProposal} – will inevitably be confronted with high fluxes of ions possessing kinetic energies $E_\textrm{kin} > \SI{5}{\electronvolt}$ originating from \textbeta-decay, or secondary ionization. This poses the challenge of possible deterioration of the substrate's spatial homogeneity, and thus the energy smearing of \textbeta-decay electrons is expected to alter over time. 

 After this successful first step, we plan to continue to study the mechanism of this tritiation method, by exploring different loading pressures and compositions, and applying improved analytical techniques.

In particular, we plan to utilise different graphene samples, custom-contacted by Graphenea with gold-layer pads, to avoid having to use the spring-loaded contacts employed during this work. This will eliminate the poor reproducibility of establishing electrical contact. In addition, appropriate contact patterns might allow for measurement options, apart from van der Pauw monitoring, using the graphene sample in a sensory capacity, like in the form of graphene field-effect transistor (gFET) sensors, which are found in an increased number of applications.\cite{Szunerits2023-pw}

 In the longer term, we intend to redesign our loading cell in such a way that access for additional monitoring tools is provided, including potentially \textit{in situ} Raman spectroscopy with spatial resolution.

Finally, we aim at investigating the applicability of the tritium–graphene system in tritium processing, such as, e.g., isotope separation. \cite{LozadaHidalgo.2017Scalableandefficient} ⁠


\bibliographystyle{unsrtnat}

\section*{Author contributions}
Conceptualization – M.S. and H.H.T. formulated the ideas for this research programme, and its goals and directions. Formal Analysis – G.Z. carried out the majority of the data analysis and visualization of data. Funding acquisition \& Project administration – M.S. and B.B. administered the overall project and secured its finances. Investigation – G.Z. and D.D.B. carried out the bulk of the experimental work, assisted in parts of the project by P.W. (Van der Pauw setup and measurements); M.A. (initial proof-of-principle work on hydrogenation of graphene, using Van der Pauw sensing); and A.L. (contacting of graphene for Van der Pauw measurements). Methodology – G.Z., D.D.B. and M.S. developed the ideas for the series of complementary measurement methodologies to reach the intended goals. Resources – S.N. was in charge of the loading-stage chamber design and, gas and sample handling in the TriHyDe facility, and N.T. was responsible for the sample heating and activity determination procedures. Software – G.Z. and J.D. developed specific software scripts to evaluate and display Raman raster scan maps, in association with our data acquisition and evaluation soft-ware suite. Supervision – H.H.T., M.S. and K.V. were responsible for overall running of the experiments, and the supervision of the research students. Writing (original draft) – G.Z., M.S. and H.H.T. prepared the draft concept for this publication, and wrote the initial manuscript. Writing (review \& editing) – all authors contributed to revising and editing of the manuscript.

\section*{Conflicts of interest}
There are no conflicts to declare. 

\section*{Acknowledgements}
G.Z. acknowledges his PhD fellowship, provided by the Karlsruhe School of Elementary Particle and Astroparticle Physics: Science and Technology (KSETA); and D.D.B. gratefully acknowledges a grant contribution for a research visit, provided by the KIT Center Elementary Particle and Astroparticle Physics (KCETA). General financial support for this CRM / graphene project has been provided by KCETA. M.S. and K.V. acknowledge seed funding by the KIT Future Fields programme through the `ELECTRON' project. Finally, we would like to thank L. Schlüter for diligently proof-reading our manuscript. 


\section*{Availability of data and materials}
The sheet resistance measurement and the individual Raman spectra are available via the corresponding publication.
Any other data is available from the corresponding author on reasonable request.


\bibliography{bmc_article}      




\end{document}